\documentclass{iaus}
\usepackage{graphicx}

\title[short title of paper] 
{Stellar population analysis of two ellipticals}

\author[short author list]   
{Andr\'e de C. Milone$^1$ %
Miriani Pastoriza$^2$ \and Mauro Rickes$^2$}

\affiliation{
$^1$Divis\~ao de Astrof\'\i sica, Instituto Nacional de Pesquisas Espaciais, Brazil
\break email: acmilone@das.inpe.br \\[\affilskip]
$^2$Instituto de F\'\i sica, Universidade Federal do Rio Grande do Sul, Brazil}

\pubyear{2007}
\volume{241}  
\pagerange{1--2}
\date{2007 Jan 26th and in revised form ??}
\jname{Stellar populations as building blocks of galaxies}
\editors{A. Vazdekis \& R. Peletier, eds.}
\begin{document}

\maketitle

\begin{abstract}
The spatial distributions of
the mean luminosity-weighted stellar age, metallicity, and $\alpha$/Fe ratio
along both photometric axes of two nearby elliptical galaxies
have been obtained using Lick index measurements on long slit spectra
in order to reconstruct the star formation history
in their kinematically distinct subsystems.
Lick indexes were compared with those of single-aged stellar population (SSP) models.
A population synthesis method was also applied
in order to help disentangling the age-metallicity degeneracy.
The stars characteristics are associated with their kinematics:
they are older and $\alpha$-enhanced in the not rotating bulge of NGC\,1052
and counter rotating core of NGC\,7796, while
they show a strong spread of $\alpha$/Fe and age along the rotating disk of NGC\,1052
and an outwards radial decreasing of them outside the core of NGC\,7796.

\keywords{
Galaxies: elliptical and lenticular, cD,
galaxies: stellar content,
galaxies: formation.}
\end{abstract}

\firstsection 

\section{Introduction}
The star formation history inside an early-type galaxy is determinate by
its formation process (merging, accretion, monolithical collapse or other).
Specifically, the stellar metallicity and age radial gradients
are dependent on the galaxy merging history
(\cite[Kobayashi 2004]{Kob04}).
Moreover, the stellar population parameters
like the age and metallicity
might be correlated with the stellar kinematics.

In this context, we have studied two distinct ellipticals
with intermediate stellar masses ($\sim$10$^{11} $\,$M_{\odot}$)
belonging to low density regions of the local Universe:
the Liner prototype E4 NGC\,1052 ($z$=0.00504, $M_{B}$=-20.50),
which belongs to a loose group and has a stellar rotating disk (i$\simeq$0$^{\circ}$),
and the E1 NGC\,7796 ($z$=0.01097, $M_{B}$=-20.79) of the field,
which shows a kinematically decoupled core (KDC).

\section{Observations}
Long slit spectroscopic observations along both photometric axes
($\lambda\lambda$4320-6360\AA, $R\simeq$1800, 2.01 \AA/pix, slit=2.08''$\times$230'')
were carried on the OPD/LNA 1.60m telescope,
providing good quality spectra up to quite 1 $r_{eff}$.
The linear spatial scales were 
111\,pc/pix and 213 pc/pix for
NGC\,1052 and NGC\,7796, respectively ($h_{0}$=0.75).

The radial profiles of the line-of-sight $\sigma_{v}$
and the line-of-sight rotational velocity curves
were satisfactorily compared with other studies.
We have confirmed the presence of a stellar rotating disk (major axis) and 
a not rotating bulge in NGC\,1052.
The stellar counter rotating core of NGC\,7796 was detected as well.

The Lick indexes of $Fe4383$ to $Na\,D$ were measured on the aperture spectra
and properly calibrated on Lick System.
Their radial gradients along both axes were computed
and the central values of some of them agree with the literature ones.
For NGC\,1052,
the $Mg\,b$, $Mg_{1}$ and $Mg_{2}$ were corrected due to the effect of emission lines
(\cite [Goudfrooij \& Emsellem 1996]{GE96}),
while $Fe5015$ and $H\beta$ were excluded from the analysis.

\section{Methods: comparisons with SSP models and population synthesis}
Firstly,
the analysis of Lick index gradients suggests a possible radial dependency for 
the Mg/Fe abundance ratio for both galaxies (and maybe for the C and N).

We have compared our Lick indexes
with the predictions of the single-aged stellar population models of
\cite [Thomas \etal\ (2003)] {TMB03}
that take the influence of abundance variations on them into account.
We have also performed a stellar population synthesis for each extracted spectrum
applying the method of
\cite [Bica (1988)] {B88}.
The relationship between the [Mg/Fe] and star formation time scale of
\cite [Thomas \etal\ (2005)] {Thomasetal05}
was employed as well.

From the SSP comparisons, we have obtained
that the stellar populations of the bulge of NGC\,1052
have [$\alpha$/Fe]$\simeq$+0.2 dex and [Z/Z$_{\odot}$]$\simeq$+0.35 dex.
Along its disk, there is a strong spread of the Mg/Fe ratio
(0. $\lesssim$ [$\alpha$/Fe] $\lesssim$ +0.5 dex) associated
with a possible outwards radial decreasing of the global metallicity to the solar value.
In the core of NGC\,7796, the populations
have nearly [$\alpha$/Fe]=+0.45 dex, [Z/Z$_{\odot}$]=+0.35 dex and 12 Gyr,
while there is an outwards radial decreasing of the Mg/Fe ratio to the solar value
associated with a possible decreasing of the global metallicity and age.

The results of the population synthesis indicate that
the nucleus of both ellipticals is dominate by old metal rich stars
($\sim$13 Gyr, $\sim$Z$_{\odot}$).
The populations are more homogeneous in the bulge of NGC\,1052 than along its disk,
where there is an outwards radial decreasing of the older-richer components
together with a respective rising contribution of the younger-metal poor ones.
The results for NGC\,7796 are analogous:
older ages and higher metallicities in the nucleus
but with similar radial behavior of age and Z along both axes.

\section{Conclusions: star formation history and chemical enrichment}
In the observed regions of both ellipticals,
the $\alpha$-enhancement is not homogeneous:
there is a monothonic radial dependency in NGC\,7796.
The global metallicity has an outwards decreasing, while
the iron abundance is nearly constant or increases outwards.
The age shows a strong spatial dispersion possibly connected to the $\alpha$/Fe spread.
The stellar populations are associated with their kinematical properties:
they are older and very $\alpha$-enhanced in the not rotating bulge of NGC\,1052
and the KDC of NGC\,7796, while
there is strong dispersion of $\alpha$/Fe and age
along the rotating disk of NGC\,1052 and 
an outwards radial decreasing of them in NGC\,7796.

Therefore, the bulk of the stars in the bulge of NGC\,1052 and KDC of NGC\,7796
was formed in an ancient short episode
providing an efficient chemical enrichment by SN-II, while
in the NGC\,1052 disk and outer parts of NGC\,7796
the star formation occurred later with larger temporal scales,
having made the enrichment by SN-Ia important.
Specifically, for NGC\,7796 an inside-out formation is plausible, while
a merging episode with a drawn out star formation is more acceptable for NGC\,1052.

\begin{acknowledgments}

A. Milone acknowledges FAPESP by the received financial aid (process n. 06/05029-3).
\end{acknowledgments}

\end{document}